\documentclass[11pt]{article}
\usepackage{moriond,epsfig}

\newcommand{\PO}{\rm l \! P }

\newcommand{\xpom}{x_{\PO} }

\def\be{\begin{equation}}
\def\ee{\end{equation}}
\def\bea{\begin{eqnarray}}
\def\eea{\end{eqnarray}}

\begin{document}
\vspace*{4cm}
\title{DIFFRACTION AT HERA AND IMPLICATIONS FOR TEVATRON AND LHC}

\author{ L. SCHOEFFEL }

\address{CEA Saclay, DAPNIA-SPP,\\
91191 Gif-sur-Yvette Cedex, France}

\maketitle\abstracts{
We describe new QCD fits to diffractive cross sections measured at HERA
and we use the parton densities derived from these fits to predict the shape of the dijet mass
fraction in double Pomeron exchange at the Tevatron. We discuss the existence of exclusive events in this
dijet channel and some prospects are given for the LHC.
}

\section{Diffraction at HERA}

One of the most important experimental results from the DESY $ep$ collider HERA
is the observation of a significant fraction (around $15\%$) of diffractive events
in deep inelastic scattering (DIS) with
large rapidity gap between  the scattered proton, which remains intact,
and the rest of the final system \cite{pdfs}.
In the standard QCD description
of DIS,  such events are not expected in such an abundance since large
gaps are exponentially suppressed due to color strings formed between
the proton remnant and scattered  partons. For diffractive events, however,
a color neutral cluster of partons fragments independently of the scattered proton.
The theoretical description of diffractive events is a
real challenge since it must combine perturbative QCD effect of hard scattering with
nonperturbative phenomenon of rapidity gap  formation. 

There are various  
interpretations of this phenomenon, but a very appealing one relies upon a partonic interpretation 
of  the structure of the Pomeron \cite{ingelman}. 
It is defined
in the presence of a hard scale, the photon virtuality $Q^2$ or  jet transverse momentum,  which
allows to apply perturbative QCD. Soft diffraction, when such a scale is missing, is outside the scope of the model
but can be described in  the context of Regge pole phenomenology. This phenomenology, however, 
turns out to be quite useful in the description of a soft part of hard diffraction, responsible for the rapidity gap formation.
 In fact, it is possible to nicely describe the 
diffractive cross-section data from HERA
 by a QCD DGLAP evolution of  
parton distributions in the Pomeron combined with  a Regge parametrisation of flux factors 
describing  the probability of finding a Pomeron state in the 
proton \cite{pdfs}.  
It follows exactly the same procedure than for standard DIS except that the diffractive distributions are 
related to the Pomeron, whose flux factor is factorised and parametrised as a function of $\xpom$,
the momentum fraction lost by the proton. 

\begin{figure}
\centerline{\includegraphics[width=0.5\columnwidth]{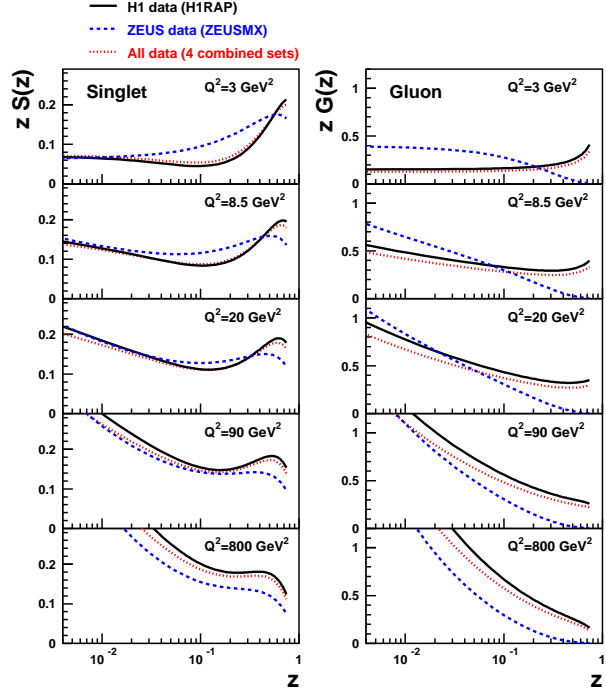}}
\caption{Singlet and gluon distributions 
of the Pomeron as a function of $z$, the fractional momentum of the
Pomeron carried by the struck parton, derived from QCD fits on H1RAP data 
alone, ZEUSMX data alone or the four data sets together. 
The parton densities are normalised to represent 
$\xpom$ times the true parton densities multiplied by the flux factor at
$\xpom = 0.003$. }
\label{gluon}
\end{figure}

Sets of diffractive parton distribution functions (dPDFs) are  shown in figure ~\ref{gluon}.
In the infinite momentum frame, the dPDFs have an interpretation
of conditional probabilities to find a parton in the proton
with the momentum fraction $x=\beta \xpom$, where
$\beta$ denotes the 
fraction of the particular parton in the Pomeron.
The gluons dominate the
diffractive exchange and carry approximately 70 \% of the momentum. 
While the quark densities are found to be relatively close for H1 and
ZEUS experiments, the gluon density differs by more than a factor 2. New preliminary data from
ZEUS reduce this discrepancy. 

In the following, we will use the QCD fits
to the H1 data to compare with the dijet mass fractions measured by the CDF
collaboration in double Pomeron exchange  and we discuss the possible evidence for 
exclusive events in this context.

\section{Diffraction at Tevatron and LHC}

The difference between diffraction at HERA and at the Tevatron is that
diffraction at the Tevatron can occur not only on either $p$ or $\bar{p}$ side as at 
HERA, but also on both sides. The former case is called single diffraction
whereas the other one double Pomeron exchange. In the same way
as we have defined the kinematical variables $\xpom$ and $\beta$ at HERA, we define $\xi_{1,2}$
as the proton fractional momentum loss (or as the $p$ or
$\bar{p}$ momentum fraction carried by the Pomeron), and $\beta_{1,2}$, the fraction of the
Pomeron momentum carried by the interacting parton. The produced diffractive
mass is equal to $M^2= s \xi_1 $ for single diffractive events and to
$M^2= s \xi_1 \xi_2$ for double Pomeron exchange,
where $\sqrt{s}$ is the energy of the reaction in the center of mass frame. The size of the rapidity gap
is of the order of $\Delta \eta \sim \log 1/ \xi_{1,2}$.

\begin{figure}
\begin{center}
\epsfig{figure=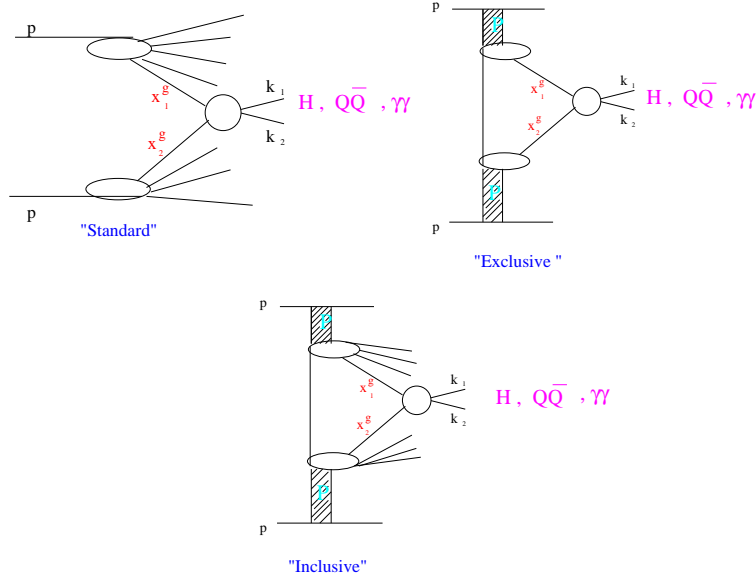,height=3.in}
\end{center}
\caption{\it Scheme of non diffractive, inclusive double Pomeron exchange and 
 exclusive events at the Tevatron or LHC}
\label{ini}
\end{figure}

It has been shown that the dPDFs of HERA can not be used directly to make predictions
at the Tevatron. Indeed, factorisation does not hold and a gap survival probability of a few \% has to be considered.
It corresponds to the
probability that there is no soft additional interaction or in other words that
the event remains diffractive. 
Knowing the presence of this essential factor, we can 
discuss the case of the double Pomeron exchange at the Tevatron.
A schematic view of non diffractive, inclusive double Pomeron exchange and 
exclusive diffractive events at the Tevatron or the LHC is displayed
in figure~\ref{ini}. The upper left plot shows the "standard" non diffractive events
where the Higgs boson, the dijet or diphotons are produced directly by a 
coupling to the proton associated with proton remnants. The bottom plot displays
the standard diffractive double Pomeron exchange (DPE) where the protons remain
intact after interaction and the total available energy is used to produce the
heavy object and the Pomeron remnants.
There may be a third class of processes displayed in
the upper right figure, namely the exclusive diffractive production. 
Exclusive events allow a precise reconstruction of the
mass and kinematical properties  of the central object
using the central detector or even more precisely  using very forward detectors
installed far downstream from the interaction point \cite{royon}. 
As mentioned above,
the mass of the produced object can be computed using
roman pot detectors and tagged protons,
$
M = \sqrt{s \xi_1 \xi_2}
$,
where  $\xi_{1,2}$ represent the fractions of energy losses for both protons. 
We see immediately the advantage of those processes : we can benefit from
the good roman pot resolution on $\xi_{1,2}$ to get a good resolution on mass. Therefore, it is
 possible to measure the mass and the kinematical properties of the 
produced object and use this information to increase the signal over background
ratio by reducing the mass window of measurement \cite{royon}. 

If such exclusive processes exist in DPE, 
the most appealing is certainly the Higgs boson
production through this channel at the LHC \cite{royon}. 
It cannot be observed at the Tevatron
due to the low production cross section, but one 
can use present measurements at the Tevatron to investigate any evidence for the existence of exclusive 
production in DPE.

\section{Dijet mass fraction at the Tevatron}

\begin{figure}[h]
\begin{center}
\includegraphics[totalheight=7cm]{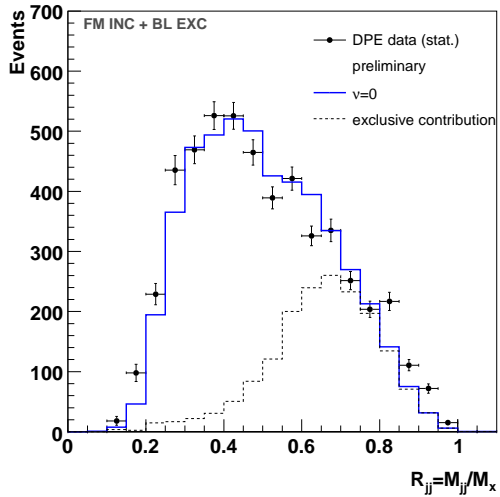}
\caption{Dijet mass fraction for jets $p_T>10\,\mathrm{GeV}$. 
The data are compared to the sum of inclusive and
exclusive predictions. The dPDFs derived from H1 data have been used together
with the survival gap probability measured with single diffractive events at Tevatron.}
\label{dijetmass}
\end{center}
\end{figure}

The CDF collaboration has measured the so-called dijet mass fraction (DMF) in dijet
events when the antiproton is tagged in the roman pot detectors
 and when there is a rapidity gap on the proton side to ensure that the
event corresponds to a double Pomeron exchange \cite{royon}. 
The measured observable  $R_{JJ}$ is defined as the ratio of the mass carried by the two jets divided by the total
diffractive mass. 
The DMF turns out to be a very appropriate observable for identifying the exclusive 
production, which would manifest itself as an excess of the 
events towards $R_{JJ}\sim 1$. 
Indeed, for exclusive events, the dijet mass is essentially equal
to the mass of the central system because no Pomeron remnant is present.
Then, for exclusive events, the
DMF is 1 at generator level and can be smeared
out towards lower values taking into account the detector resolutions.
The advantage of DMF is that one can focus on the shape of the distribution. The observation 
of exclusive events does not rely on the overall normalization which might be strongly dependent on
the detector simulation and acceptance of the roman pot detector.
Results are shown
in figure~\ref{dijetmass} with Monte-Carlo expectations calculated using DPEMC \cite{dpemc}. 
Indeed,
we see a clear deficit of events towards high values of the DMF, 
where exclusive events are supposed to occur. 
In figure~\ref{dijetmass}, a specific model describing exclusive events \cite{bl} is also added to
the inclusive prediction and we obtain a good agreement between data and the sum of MC expectations \cite{dpemc}.
It is a first evidence that exclusive events could contribute at the Tevatron \cite{royon,olda}.

\section{Conclusions}

We have discussed a first evidence for the existence of exlusive events in double Pomeron exchange at the Tevatron.
If such events can be also observed at the LHC, it would be possible to produce a Higgs boson as well as of a dijet system
regarding the cross section values accessible at the LHC.
First, a direct precise
determination of the gluon density in the Pomeron through the measurement of the
diffractive dijet cross section at the Tevatron and the LHC would be necessary
if one wants to prove the existence of exclusive events in the dijet channel.
In particular, a Tevatron or LHC diffractive gluon density could be extracted 
including {\it de facto} the survival gap probability.
Then, the great benefit of exclusive events concerns the precise reconstruction of the
mass of the central object,
using roman pot detectors
installed far downstream from the interaction point \cite{royon}. It gives the opportunity to
work with a favorable  signal/background ratio compared to standard Higgs searches 
with a mass below  150 GeV.

\end{document}